\documentclass[aps,prl,twocolumn,showpacs,superscriptaddress,groupedaddress,10pt]{revtex4-1}  
\usepackage{bm}
\usepackage{graphicx}  
\usepackage{float}
\usepackage{dcolumn}   
\usepackage{amssymb}   
\usepackage{amsmath}
\usepackage{amsfonts}
\usepackage[version=4]{mhchem}
\usepackage{siunitx}
\usepackage{xr} 
\usepackage{setspace}
\usepackage{hyperref}
\begin{document}

\title{Computing Solvation Shell Dynamics and Energetics in Electron Transfer Reactions via Molecular Dynamics Simulations}

\author{Zhenyu Wang}
\affiliation{Department of Computational Materials Design, Max Planck  Institute for Sustainable Materials, Max-Planck-Straße 1, D-40237 D{\"u}sseldorf, Germany}
\author{Mira Todorova}
\affiliation{Department of Computational Materials Design, Max Planck Institute for Sustainable Materials, Max-Planck-Straße 1, D-40237 D{\"u}sseldorf, Germany}
\email{m.todorova@mpie.de}
\author{Christoph Freysoldt}
\affiliation{Department of Computational Materials Design, Max Planck Institute for Sustainable Materials, Max-Planck-Straße 1, D-40237 D{\"u}sseldorf, Germany}
\author{J\"{o}rg Neugebauer}
\affiliation{Department of Computational Materials Design, Max Planck Institute for Sustainable Materials, Max-Planck-Straße 1, D-40237 D\"{u}sseldorf, Germany}

\begin{abstract}
Marcus theory is fundamental to describing electron transfer reactions and quantifying their rates, effectively representing the energy surface associated with an electron transfer from the reactant to the product ionic state via parabolas within a reaction coordinate diagram. Here, we present an intuitive and computationally efficient generalized reaction coordinate amenable to molecular dynamics simulations. By utilizing the nuclear charge of the ion, we are able to quantify in a targeted approach changes in the ion's solvation shell, thereby efficiently obtaining the free energy profile associated with the electron transfer, the transition state geometry and the evolution of the water network. 
\end{abstract}

\maketitle

Thermally activated electron transfer (ET) reactions are central in electrochemistry and relevant in other fields, such as physics, chemistry, biology or metallurgy. In "inner-sphere" ET reactions - which are not the topic of this work - the electron transfer is associated with a change in the position of the nucleus and occurs gradually via the formation of a (possibly transient) bond. It is then sufficient to investigate the transition state(s) as saddle point(s) of the Born-Oppenheimer potential energy surface of a single electronic state (usually the ground state) using standard transition state theory. However, if the transfer involves a sudden transition between two electronic states without any movement of the atoms, it results from thermal fluctuations which give rise to favorable configurations that allow such rapid transitions. Typically, this requires the two electronic states to be  energetically degenerate to fulfill energy conservation. 
The important role that ion solvations plays in such "outer-sphere" reactions has been recognized and exploited in Marcus theory, which has paved the way to calculate the associated Gibbs free energy of activation \cite{Marcus1956, Marcus1957, Marcus1957a, Marcus1964, Marcus1985, Marcus1993}. In Marcus theory, it is assumed that the relevant solvent reorganization that determines the relative energetic stability of the two electronic states, can be projected onto a single effective reaction coordinate $\lambda$ and the energies of these two states are typically assumed to be parabolic in $\lambda$. Their crossing point then plays the role of the transition “state” - which in fact is not a point, but a distinct sub-volume of the high-dimensional configurational space.

To turn the Marcus concept into a predictive theory for a given reaction via atomistic simulations, it is necessary to find a suitable generalized reaction coordinate and to be able to sample the thermally accessible phase space associated with each point on it. This is particularly challenging for ions in water where the geometry of the solvation shell is highly dynamic and its characteristic features, such as the average coordination number or the orientation of the solvent molecules etc. are intimately coupled with the total charge \cite{Kopper2001}. For a given charge, such configurations can of course be readily sampled with molecular dynamics (MD) calculations, with the initial (reactant) and final (product) state being obvious choices. This was recognized by Warshel \cite{Warschel1982, Warschel1987, Warschel1990} who pioneered the use of MD calculations to generate free energy curves of electron transfer reactions using a  free energy perturbation method. Utilizing umbrella sampling, the respective MD calculations run on a potential mixing the initial and final states by different ratios in a linear fashion \cite{Warschel1982, Warschel1990}. This approach has since become the foundation for studies focusing on different aspects of electron transfer reactions using empirical potentials \cite{Kuharski1988, Bader1990, Maroncelli1988, Carter1989, Yelle1997, Kopper2001}. Sprik and co-workers \cite{Sprik2002, Sprik2004, Sprik2005, Sprik2006} extended it to ab-initio MD calculations. Using a grand canonical MD \cite{Sprik2002} approach, in which the forces of the oxidized and reduced state potential energy surfaces are mixed and a constant bias  corresponding to the electron chemical potential $\mu_e$ is included in the weighting term to shift them with respect to each other, they focused on half-cell reactions \cite{Sprik2004}, which are  better suited to use in an ab-initio context. Arguing, that size effects largely cancel out when considering the whole reaction due to a similarity in the long-range effects for reactants and products \cite{Sprik2004}, they did  not apply corrections to remove spurious interactions arising from the presence of the homogeneous background charge in the resulting charged supercells. 

The present study introduces an alternative quantitative framework to study electron transfer reaction suitable for calculations using empirial or machine learning potentials, as well an ab initio approach. This approach utilizes pseudo-atoms \cite {Shiraishi1990} and allows targeted access to the metastable branches of the Marcus parabola away from the minimum, providing information about the evolution of the solvation shell and the geometry of the transition state. It is also suitable to studying electron transfer reactions at electrochemical solid/liquid interfaces, as we discuss in a forthcoming publication  \cite{Wang2025}. Here, we apply this framework to study the  prototypical oxidation/reduction electron transfer reaction of a solvated iron-ion (Fe$^{2+} \leftrightarrow $ Fe$^{3+}$) discussed by Marcus \cite{Marcus1993}, as well as the solvation of a Na-ion (Na$^0 \leftrightarrow$ Na$^{1+}$), which we find instructive when discussing reorganizations within the solvation shell. In the following, we describe our approach by focusing on the electron exchange between Fe-ions, in which Fe$^{2+}$ (Fe$^{3+}$) is oxidized (reduced) to Fe$^{3+}$ (Fe$^{2+}$). 

\begin{figure}[tp]
\centering
\includegraphics[width=0.50\textwidth]{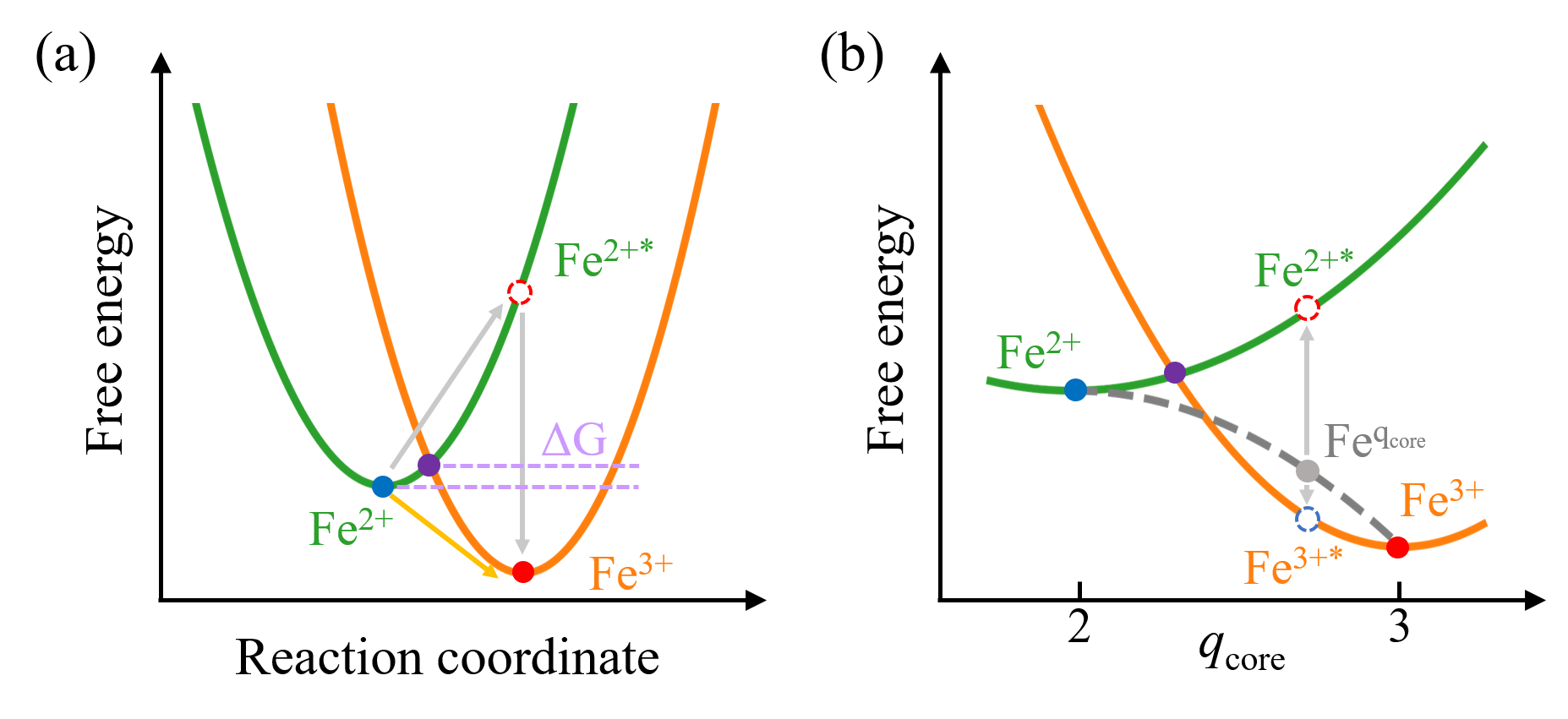}
\caption{\label{fig:MarcusParabola}
(a) Schematic free energy profile of \ce{Fe^{2+}} and \ce{Fe^{3+}} in water as a function of reaction coordinate. The ground states of \ce{Fe^{2+}} and \ce{Fe^{3+}} are marked by a red and a blue circle, respectively, while the transition state at which the charge transfer occurs is marked by a purple circle. $\Delta G$ represents the minimal energy needed for the charge transfer to happen, while the red dashed circle marks an high energy state in which \ce{Fe^{2+}} is occupying a solvation shell corresponding to the  \ce{Fe^{3+}} ground state. 
(b) Schematic free energy profile as a function of the reaction coordinate, which we will obtain from the calculations. The dashed red and green circles denote the energy of the high energy configurations Fe$^{{2+}^*}$ and Fe$^{{3+}^*}$ obtained using Eq.\,\ref{eq:UpSampling}. 
}
\end{figure}

According to Marcus, the free energy of both the initial and final state is directly proportional to the squared generalized reaction coordinate. Thus, the free energy/reaction coordinate diagram we expect to obtain would consist of two parabolas the minima of which are associated with the initial and final state of the electron transfer reaction. A corresponding schematic picture is shown ub  Fig.\,\ref{fig:MarcusParabola}\,a, where the left parabola is associated with the initial state - here, a solvated Fe$^{2+}$ (an acceptor), 
and the right parabola is associated with the final state - here, a solvated Fe$^{3+}$ (a donor).
The intersection point of the two parabolas corresponds to the transition state (also known as reorganization barrier), where the solvation shells of donor and acceptor are sufficiently similar to allow electron transfer. Accessing this configuration is, however, a rare event. It is induced by thermal fluctuations that cause reorganizations of solvent molecules which occasionally adopt (before and independently of the electron transfer) geometric arrangement which are high in energy (i.e. higher in energy compared to the equilibrium ground state) but favorable for electron transfer. These can be sampled by MD, but accessing such high energy configurations of the solvent in a targeted way is dependent on the identification of a suitable generalized reaction coordinate amenable to calculations with atomic resolution. Considering the connection Marcus theory makes between the solvent reorganization energy and the generalized reaction coordinate, observations in literature linking the ion's charge to its solvation shell \cite{Kopper2001} and the expectation, that the solvation shells of donor and acceptor are similar at the transition point where the electron is equally shared by the donor and acceptor, we chose the ion's charge as a generalized reaction coordinate. 

Varying the ionic charge ($q_\mathrm{core}$) in a continuous fashion   allows us to effectively access the high energy points along the Marcus parabola (cf. Fig.\,\ref{fig:MarcusParabola}), which provides us with  direct access to the corresponding arrangements of the water molecules in the solvation shell.  We achieve such a variation of the ion's charge ($q_\mathrm{core}$), by utilizing Shiraishi's idea of pseudo-atoms \cite{Shiraishi1990}. Using pseudo-atoms, which are neutral objects due to having the same (fractional) core and valence charge, has the advantage that our MD calculations are carried out around the ground state configuration of each ion $D$ (here, $D$ = Fe, Na or Cl) in charge state $q_\mathrm{core}$ (i.e. $D^{q_\mathrm{core}}$) we consider. 

The calculations presented in this paper are carried out using the LAMMPS package \cite{Plimpton1995, Thompson2022} with empirical potentials, but the methodology is equally well applicable in an ab initio context. Here, we create pseudo-atoms by adjusting the Coulombic repulsion term of the potential to construct ions with fractional charges ($q_\mathrm{core}$). We use cubic simulations cells of $(27 \times 27 \times 27)${\AA} containing 658 water molecules and a single ion $D^{q_{\mathrm core}}$ ($D$ = Fe, Na (or Cl \cite{SM})) with integer or fractional charge ${q_{\mathrm core}}$. The water is described at the level of a TIP3P potential \cite{Price2004} and the ion-water interactions are modeled using a Lennard-Jones (12-6) potential together with Coulombic terms capped at \SI{10.0}{\AA}. For the MD simulations we use a Langevin thermostat \cite{Schneider1978, Dunweg1991}, which ensures that the targeted temperature of \SI{300}{\kelvin} is sustained over the \SI{100}{\nano\second} long production trajectory. \footnote{Further details about the computational parameters, the performed convergence test and the minimum cell size required to ensure that we capture the relevant solvation shells and their dynamics to achieve converged results are provided in the Supplementary Information \cite{SM}.}
Using the MD calculations we calculate the formation energy of each solvated $D^{q_{\mathrm core}}$ ion \cite{Todorova2014} as: 
\begin{equation}
\label{eq:formation}
\begin{aligned}
    E^f ({\rm N}a^{q_{\rm core}}) &= 
    \langle U ({\rm D}^{q_{\rm core}}:{\rm host})\rangle - \langle U ({\rm host}) \rangle \\
    &- \sum_j \Delta n_j \cdot \mu_j + q \cdot \mu_e \quad . 
\end{aligned}
\end{equation}

This corresponds to the conventional formula for calculating formation energies of charged point defects in semiconductors \cite{Neugebauer2004, Freysoldt2014}, except for using average potential energies obtained from the MD calculations for the solvated ion $\langle U (D^{q_{\rm core}}:{\rm host})\rangle$ (here: Fe$^{2+}$, Fe$^{3+}$, Na$^0$, Na$^{+1}$ or Cl$^0$, Cl$^{-1}$) and the neat solvent $\langle U ({\rm host}) \rangle$ (here: water). Differences in the number of species $j$ between these two MD calculations are accounted for by $\Delta n_j$ and the species's chemical potential $\mu_j$ \footnote{The chemical potentials act as external reservoirs for the exchange of species or electrons, guaranteeing a grand canonical description}. $\mu_e$ stands for the chemical potential of the electron \cite{Todorova2014}, and becomes particularly relevant when potential control is applied to achieve anodic or cathodic conditions. 

\begin{figure}[tp]
\centering
\includegraphics[width=0.45\textwidth]{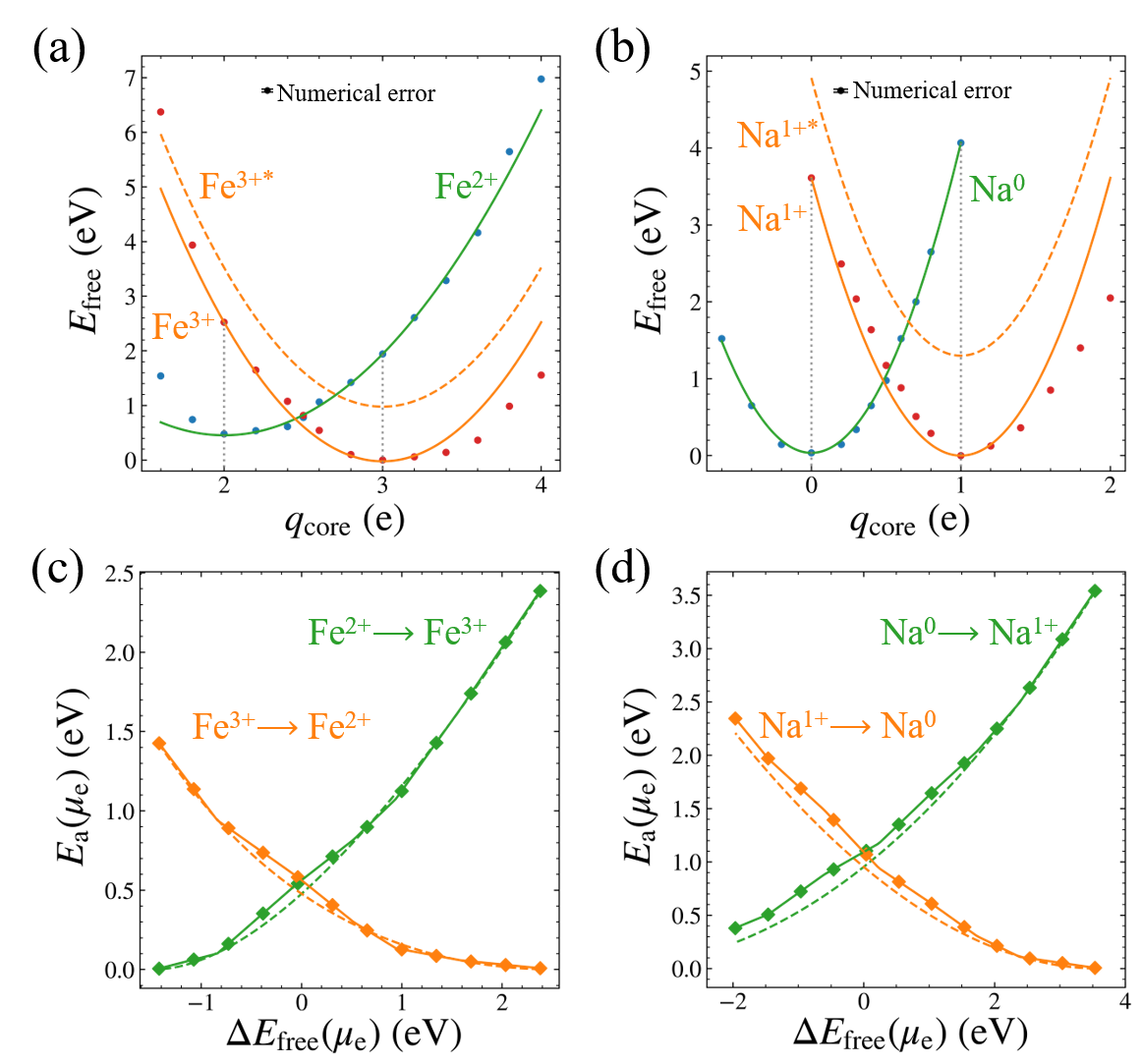}
\caption{\label{fig:EnFreeNa}
Calculated free energy profiles for (a) Fe$^{2+}$ and Fe$^{3+}$ and (b) Na$^0$ and N$^{1+}$. The actually calculated free energies obtained from the MD-runs for different $q_{\mathrm{core}}$ values are shown as points. Error bars are omitted for clarity, since the standard error of the mean (SEM) for each point is less than 25\,meV. Parabolas fitted using the ground state (i.e. minimum  Fe$^{2+}$ or Fe$^{3+}$ energy) and its corresponding high energy vertical excitation state (i.e. Fe$^{{2+}*}$ or Fe$^{{3+}*}$ are shown as solid lines and clearly reveal deviations from the calculated points. Applying, e.g., positive bias $\mu_e > 0$ shifts the Fe$^{3+}$ (Na$^{1+}$) parabolas upward, as shown by the dashed orange curves in (a) and (b).\\
The free energy barrier for electron transfer and how it changes as the electron chemical potential $\mu_e$ is varied (i.e. application of bias) is shown in (c) for Fe$^{2+} \rightleftharpoons$ Fe$^{3+}$ and in (d) for  Na$^{0} \rightleftharpoons$ Na$^{1+}$. The dashed lines depict the free energy barrier variation obtained by shifting the parabolas fitted in (a) for Fe$^{2+}$ and Fe$^{3+}$ (respectively (b) for Na$^0$ and Na$^{1+}$) with respect to each other by a bias potential $\mu_e$. Connecting the MD-calculated points in (a) (respectively (b)) by linear segment and then shift the ensuing curves by $\mu_e$ with respect to each other, results in  the solid lines. The squares each of the two lines separate regions where the slopes of the intersecting linear segments remain unchanged. }
\end{figure}

The resulting formation energies for the solvated Fe$^{q_\mathrm{core}}$ ions give raise to the gray dashed curve in Fig.~\ref{fig:MarcusParabola}b, which interpolates smoothly the energy between the two energy minima (i.e. Fe$^{2+}$ and Fe$^{3+}$) and covers all explicitly calculated points along the reaction coordinate. Particularly interesting is, that from the MD calculations we obtain a representative distribution of the solvation structures associated with high energy states along the Marcus parabola, which is very useful, because it allows us to extract from each Fe$^{q_{\mathrm{core}}}$ trajectory $N$ uncorrelated snapshots. Using these snapshots, we perform two additional static calculations where we substitute the charge $q_{\mathrm core}$ of Fe$^{q_{\mathrm core}}$ once by $2+$ and once by $3+$, as indicated by the vertical gray arrows in Fig.\,\ref{fig:MarcusParabola}\,b, by which calculation we obtain the high energy $Fe^{{2+}*}$ and $Fe^{{3+}*}$ points of the Marcus parabola. 

We note, that each of these static calculations is for a charged system, whenever we consider charges $q_{\mathrm{core}} \ne$ 2+ or 3+. The reason is, that the solvation shell obtained by the MD calculation for the ion $D^{q_{\mathrm core}}$ effectively screens only this charge. Consequently, when we exchange the Fe$^{q_{\mathrm{core}}}$ ion with a Fe$^{2+}$ or Fe$^{3+}$ ion, part of their charge remains unscreened - the Fe$^{{2+}*}$ ion is over-screened by a charge $q_\mathrm{core}$, whereas the Fe$^{{3+}*}$ ion is under-screening by a charge ($1 - q_\mathrm{core}$). These unscreened charge needs to be corrected for, as it gives rise to  spurious charge interactions with its periodic images and with the  homogeneuos background automatically present in any periodic DFT calculations for a charged supercell. Fortunately, calculations for charged point defects are common for semiconductors and suitable correction schemes exist \cite{Freysoldt2014}. Using the scheme suggested by Freysoldt et al. \cite{Freysoldt2009} we obtain each each high energy configuration $i$ we consider a correction term $\Delta E$\,\footnote{A detailed discussion about the charge correction is provided in the Supporting Information \cite{SM}}    

\begin{equation}
    \label{eq:UpSampling}
    \langle \Delta E ({\rm Fe}^{q^*} - {\rm Fe}^{q_{\rm core}})\rangle = 
    \frac{1}{N} \sum_{i=0}^N 
    \langle U_i ({\rm Fe}^{q^*}) - U_i ({\rm Fe}^{q_{\rm core}}) \rangle
\end{equation}
\\
which we use to correct the potential energy $U_i ({\rm Fe}^{q^*})$ and add to the calculated formation energy (eq.\,\ref{eq:formation}) to obtain the energies of the donor and acceptor state compatible with the current solvation shell (orange and green line in Fig.\,\ref{fig:MarcusParabola}\,b).

The final contribution we need to account for is entropy and we calculate it using thermodynamic integration \cite{}. This provides us with the free energy difference between the two ground states (Fe$^{2+}$ and Fe$^{3+}$) and allows us to align them by integrating the average of their potential energy difference over the reaction coordinate ($q_{\mathrm{core}}$) from 2$e^-$ to 3$e^-$:

\begin{equation}
\label{eq:ThermoDy}
\begin{aligned}
&\Delta F_{\mathrm{shift}}(\mathrm{Fe^{2+}} \rightarrow \mathrm{Fe^{3+}}) = F (\mathrm{Fe^{3+}}) - F (\mathrm{Fe^{2+}}) \\
&=\int_{2}^{3} \frac{\partial F (q_{\mathrm{core}})}{\partial q_{\mathrm{core}}} \, dq_{\mathrm{core}} = \\
&= \int_{2}^{3} {\left\langle U (\mathrm{Fe^{3+}}) - U (\mathrm{Fe^{2+}}) \right\rangle}_{q_{\mathrm{core}}} \, dq_{\mathrm{core}}\:.  
\end{aligned}
\end{equation}
\\
The resulting free energies profiles we calculate for Fe$^{2+}$, Fe$^{3+}$, Na$^0$ and Na$^{1+}$ are seen in Fig.\,\ref{fig:EnFreeNa}\,a and b. 

\begin{figure}[htp]
\centering
\includegraphics[width=0.45\textwidth]{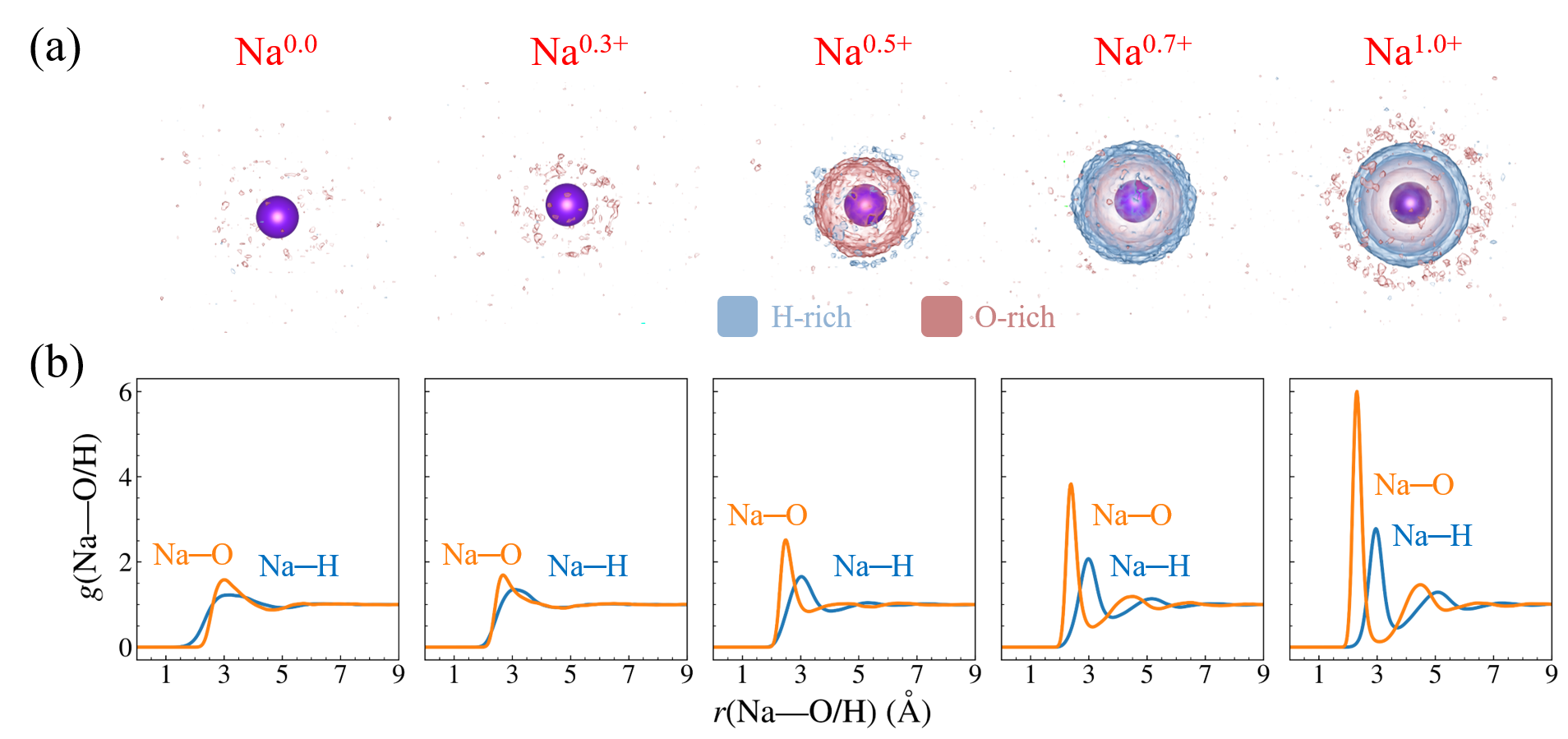}
\includegraphics[width=0.15\textwidth]{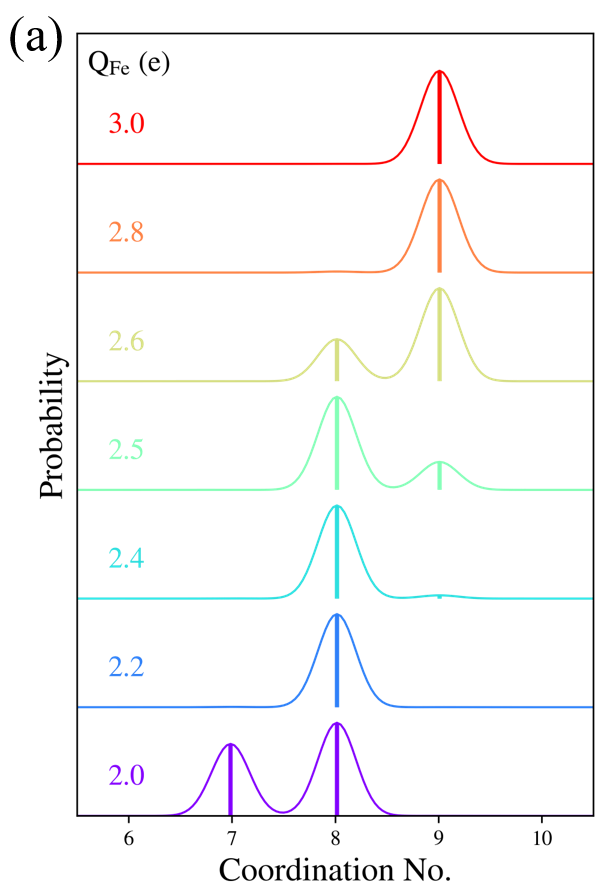}
\caption{\label{fig:SolRdfNa}
(a) Evolution of Na-ion solvation shell as a function of the charge state of the ion ($q_{\rm core}$) with positive (H) and negative (O) charges in the solvation shell shown in blue and red, respectively. 
(b) The corresponding radial distribution functions (RDFs) for Na-O and Na-H as a function of the charge state  $q_{\rm core}$ of the Na ion. 
(c) Evolvement of the coordination number of Fe to water molecules within the first solvation shell of a Fe$^q$ ion, as it charge changes from $2^+$ to $3^+$. A similar plot for Na is shown in the supplemental material \cite{SM}.
}
\end{figure}

Using two points - the energy minimum corresponding to the ground state solvation configuration of Fe$^{2^+}$ (Fe$^{3^+}$) and its vertical transition high energy equivalent Fe$^{{2^+}^*}$ (Fe$^{{3^+}^*}$), where a Fe$^{2^+}$ ion resides within the solvation shell of a Fe$^{3^+}$ ion (and vice versa) - we fit parabolas to the calculated ion formation free energies. Applying this procedure to both Fe and Na we observe in each case clear deviations from a parabolic behavior. These is seen in the vicinity of the intersection of the two curves, which signify the region where the  electron transfer will occur, and in the inability to fit one unique parabola which describes equally well points on the right and left side of  the ground state. 

The intersection point of the two free energy curves corresponds to the barrier configuration. In the absence of an applied bias (i.e. $\mu_e =0$\,eV) we find, that the charge state at which the electron transfer occurs is slightly shifted from the haft point between the two ground states (i.e. the transition between Fe$^{2^+}$ and Fe$^(3^+)$ occurs at $q_\mathrm{core} = 2.51\,e^-$ and for Na$^0$ and Na$^1+$ at $q_\mathrm{core} = 0.52\,e^-$\,\footnote{Further details on the determination of this value are found in the Supplementary Information}. 

Application of bias shifts the curves for the reduced/oxidized state of an ion with respect to each other, modifying the barrier. Accounting for such shifts in the transition point due to external influences, such as the application of electric fields, is straight forward within our approach, since the formula for the formation energy (Eq.\,\ref{eq:formation}) includes the chemical potential of the electron. A bias potential for which $\mu_e < 0$\,eV facilitating oxidation (i.e. a downward shift of the transition point, which reduces the activation barrier from left to right), while $\mu_e > 0$\,eV facilitates reduction. This is seen in Fig.\,\ref{fig:EnFreeNa}\,c and d, which show the how the free energy difference and the activation energy change relative to each other with  varying bias. When the free energy difference is $\Delta E_{\mathrm free} = 0$) the ground states (left and the right minimum for Fe$^q$, respectively Na$^q$, in Fig.\ref{fig:EnFreeNa}\,a and b) align. When the two curves intersect at one of the minima, the activation barrier $E_{\mathrm a}$ vanishes for electron transfer from the charge state corresponding to the minimum at the intersection point to the other minimum. Furthermore, we note that the observed deviations of the calculated free energy curves from a parabolic behavior skews the activation energy curve, resulting in slightly higher energies $E_a$. 
is
Because the charge is directly related to the arrangement of the water molecules within the solvation shell of the ion undergoing oxidation or reduction, using $q_\mathrm{core}$ as a reaction coordinate provides direct access to the solvation shell evolution associated with such a reaction. This is depicted in Fig.~\ref{fig:SolRdfNa}, which shows how the solvation shell of of Na evolves as the charge of the ion increases from 0 to 1. The upper panel (a) shows the spatially resolved charge density of water molecules around the ion Na$^{q_{\rm core}}$ and reveals that at ca. $q_\mathrm{core} = 1/2e^-$ the first solvation shell around the ion is fully build. After this point, the second shell starts to form and it is completely formed at full charge, i.e. for Na$^{1^+}$. This can be seen even clearer in Fig.~\ref{fig:SolRdfNa}\,b, which depicts the radial distribution function (RDF) for Na--H and Na--O. There lack of a solvation shell for a neutral Na, which just presents an exclusion volume for the water molecules, results in an overlap of the Na--H and Na--O peaks associated with the Na's nearest water neighbors. 
The gradual increase of the Na charge leads to the formation of a first solvation shell (also seen in Fig.~\ref{fig:SolRdfNa}\,a), which is completed at about $q_{\rm core}$ = 0.5$e^-$, at which point the first Na--H and N--O RDF peaks are distinct and clearly separated. After this point a second Na--O and Na--H peak starts to evolve, signifying the emergence of a second solvation shell, which is completed at full charge (i.e. Na$^{1^+}$) where these two peaks are distinctly shifted with respect to each other. We  therefore conclude, that the charge transfer occurs when the first solvation shell of the ion is completed, as this provides sufficient screening of the charge to render the charge transfer favorable. 

A similar conclusion can be drawn in the case of the Fe$^{2^+}$ to Fe$^{3^+}$ electron transfer, although the changes to the solvation shell are a bit more subtle and less easily visualized, since both ions already possess a fully build solvation shell. The change in solvation structure can be detected, by inspecting the change in coordination number of the water molecules within the first solvation shell of the Fe-ions. While Fe$^{2^+}$ is coordinated to 8 H$_2$O molecules, Fe$^{3^+}$ is coordinated to 9 water molecules. This transition from 8 to 9 H$_2$O molecules occurs again halfway in between the initial and final state, in  the range of charges associated with the transition state (cf. Fig.~\ref{fig:SolRdfNa}\,c).

In this study we presented an explicit and physically intuitive  generalized reaction coordinate to describe charge transfer in aqueous environments. The suggested $q_{\rm core}$ (ionic charge) coordinate allows to access the high energy configurations along the transition path of an electron transfer reaction in a targeted way, thereby providing an computationally efficient means to access information about the free energy and geometric changes associated with the transition state at which the electron transfer occurs. The methodology is quantitative, reproduces the relevant branches of a Marcus' parabola, directly probes the reorganization barrier and allows accessing information about changes in the solvation shell (e.g. coordination number to water molecules) during the process. Furthermore, the effect an externally applied potential would have on the transition barrier can be straightforwardly included. Altogether, the results offer a general and effective computational tool to study and predict charge transfer processes in non-crystalline systems, paving the way towards a targeted design of electrochemical systems with desired functionality.  

\begin{acknowledgments}
The authors acknowledge financial support from the European Union’s Horizon 2020 research and innovation program under the Marie Sk\l{}odowska-Curie actions (MSCA) grant agreement No. 801459-FP-RESOMUS and the RESOLV program by the Deutsche Forschungsgemeinschaft (DFG, German Research Foundation) under Germany’s Excellence Strategy-EXC 2033-390677874-RESOLV.
\end{acknowledgments}

\bibliography{main}
\end{document}